\newcommand{\Pl}{\partial}
\newcommand{\ts}{\textstyle}
\newcommand{\fder}[2]{\frac{{\ts d \/ #1}}{{\ts d\/ #2}}}
\newcommand{\fpar}[2]{\frac{{\ts \Pl \/ #1}}{{\ts \Pl \/ #2}}}
\newcommand{\nder}[3]{\frac{{\ts d^{#1} \/ #2}}{{\ts d \/ #3^{#1}}}}

\newcommand{\bee}{\begin{equation}}
\newcommand{\ene}{\end{equation}}
\newcommand{\beea}{\begin{eqnarray}}
\newcommand{\enea}{\end{eqnarray}}

\documentclass[prb,aps,twocolumn]{revtex4}
\usepackage{dcolumn}
\usepackage{graphicx}
\usepackage{latexsym}
\usepackage{amsmath}
\usepackage{amssymb}

\begin{document}
 
\title{Excitation of plasma wakefields by intense ultra-relativistic proton beam}
\author{Mithun Karmakar,$^{1}$  Bhavesh Patel,$^{1}$ Nikhil Chakrabarti,$^{2,3}$ and Sudip Sengupta$^{1,2}$}
\affiliation{$^1$ Institute for Plasma Research, Bhat, 382428 Gandhinagar, India
\\$^2$Homi Bhaba National Institute, Training School Complex, Anushakti Nagar, Mumbai 400085, India
\\$^3$Saha Institute of Nuclear Physics, 1/AF Bidhannagar, Kolkata-700064, India }
\begin{abstract}
{We report here an exact analytical travelling wave solution for nonlinear 
electron plasma wave excited by an intense ultra-relativistic proton beam. 
It brings out the underlying physics of longitudinal electric field characteristics of the 
excited wake wave formed behind the drive beam. 
The results are further supplemented by a fully relativistic particle in cell (PIC) code OSIRIS in 2D geometry. The plasma and beam parameters in the simulation are chosen in conformity with experimental work and it provides us the anticipated axial and transverse electric field profiles. 
The investigation is further extended by providing an analytical description of the wake wave excited
by  equi-spaced train of small proton bunches with the inclusion
of the non-relativistic plasma ion dynamics. Our results show that the amplitude 
of the wake field does not grow indefinitely with the increase in the number 
of proton bunches. On the contrary, it saturates to a definitive limit.}
\end{abstract}

\maketitle

\begin{section}{Introduction}
The supremacy of plasma based particle accelerators over many conventional linear accelerators is now accepted undeniably. This novel acceleration scheme not only miniaturizes the size of an accelerator but also makes it possible to produce electrons even with TeV energies.
Over the last few decades, research on plasma based acceleration processes was mainly 
focused on  the physics of wake wave excitation in a plasma driven by a highly relativistic electron beam or an intense laser pulse.\cite{rosenbluth,tajima,rosenzweig,holkundkar,leemans,joshi,kourakis1,kourakis2,chen1,chen} 
 In electron or laser driven plasma wake field accelerators, energies up to several GeV 
can be achieved.\cite{leemans,joshi,chen1,chen} However, to reach the energy frontier of the present day high energy physics 
research, it is required to accelerate a charged particle up to an energy which is in the range of several TeV. 
It takes multiple stages of acceleration to reach TeV order of energy in laser pulse or electron bunch driven
schemes which results in various technological difficulties. Moreover, the energy gain is limited by the energy carried by
electron driver which is very small ($\sim 100$ J). An alternative approach is to 
use a proton beam instead of an electron or laser beam to excite the plasma wake wave. Because of their higher energy ($\sim$ kJ) and mass, protons can drive 
wake fields over much longer plasma lengths. A proton bunch carrying 
energy of the order of $\sim $ kJ is capable of accelerating electrons in the TeV energy range in a single plasma stage.\cite{assmann,lotov2}
 Such proton beams are now routinely produced at
various proton synchrotron facilities like Large Hadron Collider (LHC, 6.5 TeV, $1.2\times 10^{11}$ 
protons, $\sim 125$ kJ) and  CERN Super Proton Synchrotron (SPS, 450 GeV, $3 \times 10^{11}$ protons, 
$\sim 20$ kJ). \cite{xia} Therefore, because of high energy content, easy availability of proton beam
and also technological viability to use it as drive beam  
to reach TeV energy regime, this new acceleration scheme becomes superior to others.

In case of negatively charged driver, background
plasma electrons are expelled to form a
bubble surrounded by a thin electron sheath. Proton beams, on the other hand, {\it ‘suck in’} the 
plasma electrons towards the propagation axis, creates oscillations which propagate at nearly the speed of light along with the proton drive beam. 
The energetically favourable bubble solution obtained using a negatively charged drive beam, however, is very difficult to achieve by using proton drive beam. Nevertheless, after the formation of wake wave, a trailing witness bunch of electrons injected externally at proper phase of the wake with sufficient energy will be trapped and accelerated by the longitudinal electric field of the wake to relativistic energies.

The concept of wake field excitation using proton beam was first realized in 2009 by Caldwell {\it et al.}\cite{caldwell3} 
They performed simulations to study the plasma wave excitation using PIC code VLPL as well as LCODE. 
A proton beam of longitudinal size 100 $\mu$m and transverse size 0.43 mm was used to excite the wake field in a
plasma of density $6\times 10^{14}$ cm$^{-3}$. With this plasma density, the plasma wavelength of the wake wave is of the order of transverse size of the beam. It was shown that proton beam with initial energy of a $\sim $ TeV can produce longitudinal wake field of amplitude $\sim$ 3 GV/m which accelerates $\sim$ 10 GeV externally injected witness beam of electrons up to an average energy of $\sim$ 0.62 TeV after traversing a distance of $\sim$ 450 m. A strong transverse field was also observed which focussed the witness electron beam and thus reduced the transverse spreading of the beam. After this initial study, several plasma physics groups got involved in exploring the physics of PDPWFA (Proton Beam Driven Plasma Wake Field Accelerator) either through  simulation, theory or by modelling experiments on it.\cite{caldwell3,caldwell2,lotov4,vieira,kumar,caldwell1,lotov2}
The AWAKE (Advanced Wakefield Experiment) project at CERN is the first experimental project worldwide which aims to  accelerate electrons in the TeV energy range by exciting plasma wake field using modulated proton beam of TeV order of energy. \cite{caldwell2,caldwell4,bracco,pepitone}
A 400 GeV/c proton beam extracted from CERN SPS is used to excite a wake field in a 10 m long plasma cell which is capable of producing a longitudinal electric field of amplitude upto several $\sim$ GV/m.
AWAKE uses plasma densities in the range 10$^{14}$ - 10$^{15}$ cm$^{-3}$ which correspond to plasma wavelength of the wake wave to be of the order of $\sim$ few mm. 
 The proton bunches available
today are much longer in size compared to the plasma wavelength. \cite{xia}
The proton beam which is available in CERN SPS with TeV energy is 12 cm long. So they are not resonant and
 excitation of strong wake field is not possible. However, a process called self modulational instability
 can cause a long proton bunch to split into a large number of micro-bunches which then efficiently excite
 the plasma wake wave. \cite{vieira,kumar,caldwell1,lotov1,schroeder1,schroeder2}
 In the recent past, the excitation mechanism of wake field by such trains of equidistant particle bunches has been discussed.\cite{lotov3} The AWAKE experiment uses such 
modulated proton beam in the wake field excitation process.

Admittedly, it is really a challenging job to develop a multidimensional theory for PDPWFA in the nonlinear regime. In our present investigation, we first present analytical results obtained in one dimension for the wake field excited by a single bunch of protons.
These analytical results are further verified by 2D OSIRIS simulation.\cite{fon,hem} Then, the field excited using a train of proton bunches is discussed.

The paper is organized as follows. In the second section, exact analytical solution is 
presented for the wake wave excited by a single proton drive bunch in an un-magnetized plasma system.
In the third section 2D PIC simulation results
are shown. In the subsequent section, we discuss the wake wave excited by a train of proton micro bunches. The main conclusion of the paper is provided in the fifth section.
   
\end{section}

 \begin{section}{Analytical solution for the wake wave excited by an ultra-relativistic proton beam}
We adopt simple fluid model to describe the wake field excitation 
in a two component un-magnetized electron-ion plasma. The excited wake wave is a longitudinal one dimensional
electrostatic electron plasma wave propagating along the direction of propagation of a highly relativistic 
proton beam. Exact analytical solution of the problem can be obtained from the 
following fluid equations coupled with the Maxwell's equations viz. continuity equation, 
the electron fluid momentum, and 
Poisson's equation:

 \beea
\fpar{n_e}{t}+\fpar{}{x}(n_e v_e)=0,
\label{a1}
\enea
 
 \beea
\fpar{{ p_e}}{t}+v_e\fpar{p_e}{x}=-e E_x,
\label{a2}
\enea
 \beea
\fpar{E_x}{x}= 4 \pi e(n_0-n_e+n_b).
\label{a3}
\enea
The proton beam is characterized by its density $n_b$ and velocity $v_b$.
The heavy plasma ions are considered to be immobile maintaining a overall charged neutrality 
in the equilibrium plasma system. The electric field is denoted by $E_x$ which is directed 
along the positive $x$ axis. All other variables have their usual meanings.

We construct a stationary wave solution of Eq.(\ref{a1})-Eq.(\ref{a3}) considering 
the propagation of a longitudinal electrostatic wave along $x$ axis assuming all the dynamical variables 
to be function of 
$\xi=k_p(v_{ph}t-x)$, a special combination of space and time. Here $k_p=\omega_p/v_{ph}$ and
$\omega_p=\sqrt{4\pi n_{0} e^2/m_e}$ with
$n_{0}$ and $v_{ph}$ being the equilibrium plasma density and the phase velocity of the longitudinal plasma 
wave respectively. We rescale the variables by introducing 
$n=n_e/n_0$, $\alpha=n_b/n_0$, $\beta=v_e/c$, $\beta_{ph}=v_{ph}/c$, $E=eE_x/m \omega_{p}v_{ph}$, $p=p_e/mc$ with $c$ being the 
velocity of light in free space. We assume a rectangular proton beam profile with the longitudinal extension given by
\beea
\alpha &=& \alpha_0 ~{\rm{for}}~ 0\le \xi \le l_b, \nonumber \\
&=&  0  ,         \mbox{otherwise}. \nonumber
\enea
where $l_b$ is the beam length. Now it is important to note the fact that the phase velocity 
of the wake wave is determined by the velocity of drive beam. Also since we are primarily interested in investigating the physics of wake wave excited by an ultra-relativistic proton beam, we work in the limit $\beta_b \rightarrow 1$ and hence $\beta_{ph} \rightarrow 1$. In the transformed co-ordinate system, the electron continuity equation [Eq.(\ref{a1})] is directly integrated to 
give 
\beea
n=\frac{1}{1-\beta}.
\label{a6}
\enea
Hence, the wake wave excitation is now determined by the following equations  
\beea
(1-\beta)\fder{E}{\xi}=(1+\alpha)\beta-\alpha, 
\label{a4}
\enea
 
\beea
(1-\beta)\fder{p}{\xi}=-E.
\label{a5}
\enea
where we have used [Eq. (\ref{a6})] in the Poisson's equation to arrive at [Eq. (\ref{a4})]. Now we proceed to find the solution of the above equations inside and outside the beam separately. 
\begin{subsection}{SOLUTION INSIDE THE BEAM }
Combining Eq.(\ref{a4}) and Eq.(\ref{a5}) and by making another  
variable transformation:
\beea
(1-\beta)\fder{}{\xi}=\fder{}{\varphi}; 
\label{a7}
\enea
we obtain

\beea
\nder{2}{p}{\varphi}=-\left[\frac{(1+\alpha_0)p-\alpha_0}{\sqrt{1+p^2}}\right]. 
\label{a8}
\enea
This equation is integrated once to obtain
\beea
\fder{p}{\varphi}=\pm \sqrt{2}[C+\alpha_0 p -(1+\alpha_0)\sqrt{1+p^2}]^{1/2},
\label{a9}
\enea
with $C$ being an integration constant.

In order to get an exact analytical solution of the above 
nonlinear first order differential equation we
use the following transformation relations
\beea
\sqrt{1+ p^2} &=&X^2-p,\nonumber\\
a^2 &=& C+[C^2-(1+2\alpha_0)]^{1/2},\nonumber\\
b^2 &=& C-[C^2-(1+2\alpha_0)]^{1/2}. \nonumber
\enea
We assume that at $\xi=0$, $p=0$ and $\fder{p}{\varphi}=0$. Therefore the constant of integration
$C$ becomes $(1+\alpha_0)$. By using all the above transformations we obtain from Eq. (\ref{a9}) 
\beea
\varphi=\int_1^X \frac{X^2 dX}{\sqrt{(a^2-X^2)(X^2-b^2)}}\nonumber\\
+\int_1^X \frac{dX}{ X^2 \sqrt{(a^2-X^2)(X^2-b^2)}}.\nonumber
\label{a10}
\enea
We integrate the above two integrals with the substitution 
\bee
\sin^2 \theta = \frac{a^2-X^2}{a^2-b^2}.\nonumber
\ene
After performing some simple algebra, we obtain
\beea
\varphi=-\left\{a+\frac{1}{a^3(1-k^2)}\right\}[E(\theta,k)]_{\theta_0}^{\theta}
+ \left[\frac{k^2\sin\theta\cos\theta}{a^3(1-k^2)}\right]_{\theta_0}^{\theta}.\nonumber
\enea
  
\begin{figure}[h!]
{\centering
{\includegraphics[width=3.4in,height=2.2in]{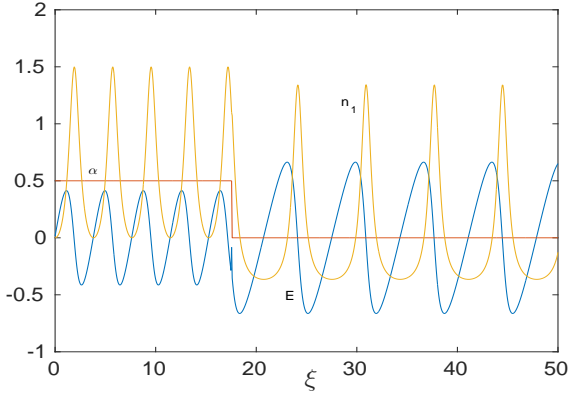}\par}}
\caption{Normalized electric field and electron density ($n_1=n-1$) in a proton 
beam driven plasma,
with beam density [$\alpha=0.5$ for $0 \le \xi \le 5.6 \pi$ and zero otherwise].}
\label{fig.1}
\end{figure}



Now $\varphi$ is related to $\xi$ in the following fashion,

\beea
 \varphi=\xi+\int\beta d\varphi .\nonumber
\label{a11}
\enea
Eventually, we can write the exact travelling wave solution for the proton beam driven plasma wake field as,
\beea
 \xi=-\frac{2}{a^3(1-k^2)}[E(\theta,k)]_{\theta_0}^{\theta}
 + \left[\frac{k^2\sin 2\theta}{a^3(1-k^2)}\right]_{\theta_0}^{\theta},\nonumber\\
\label{a12}
\enea 
where, $E(\theta,k)$ is the Elliptic Integral of second kind and\\
\beea
k^2 &=&\frac{a^2 -b^2}{a^2},\nonumber\\
\theta &=& \sin^{-1} \sqrt{\frac{a^2-X^2}{a^2-b^2}},\nonumber\\
\theta_0 &=& \sin^{-1} \sqrt{\frac{a^2-1}{a^2-b^2}}.\nonumber 
\enea

\end{subsection}

\begin{subsection}{SOLUTION OUTSIDE THE BEAM }
The solution behind the beam can be obtained by setting $\alpha=0$ in Eq.(\ref{a4}). The differential equation that we have to solve in this case is 

\beea
\nder{2}{}{\xi}\frac{1-\beta}{\sqrt{1-\beta^2}}=\frac{\beta}{1-\beta}. 
\label{aa1}
\enea
This equation is integrated with the substitution $y=\sqrt{\frac{1-\beta}{1+\beta}}$ to obtain
\beea
\frac{1}{2}\left(\fder{y}{\xi}\right)^2 +\frac{1}{2}\left(y+\frac{1}{y}\right)=\gamma_m,
\label{aa2}
\enea
where, $\gamma_m$ is the constant of integration. We can determine the value of
$\gamma_m$ by writing down the differential equation describing field characteristics inside the beam [Eq.(\ref{a4}) and Eq.(\ref{a5})] in the following form \cite{bera}
\beea
\left(\fder{y}{\xi}\right)^2 = 2(1+\alpha_0)-\frac{1}{y}-(1+2\alpha_0)y,
\label{aa3}
\enea
and by using the continuity condition of $y$ and $\fder{y}{\xi}$ at $\xi=l_b$.
Thereby, from equations Eq. (\ref{aa2}) and Eq. (\ref{aa3}), we obtain the value of $\gamma_m$ as
\beea
\gamma_m= (1+\alpha_0)+\alpha_0 y_b.
\label{aa4}
\enea
where $y_b$ is the value of $y$ at $\xi=l_b$. From the solution (inside the beam) presented 
in the earlier subsection,  we can calculate $y_b=\sqrt{\frac{1-\beta_b}{1+\beta_b}}$, 
where $\beta_b$ is the value of $\beta$ at $\xi=l_b$. Then we proceed to obtain 
the solution behind the beam in the same way as obtained by Bera ${\it et}$ ${\it al.}$ 
for the case of an electron beam driver \cite{bera}. 
The solution for the wake wave electric field behind the beam can be expressed as
\beea
E(y)=\pm \sqrt{2(1+\alpha_0)-2\alpha_0 y -(y+1/y)}.
\label{aa5}
\enea
In order to find the solution for the electric field and electron density as a function of $\xi$, 
we integrate Eq. (\ref{aa2}) to obtain
\beea
\xi =l_b +2\sqrt{b}[E(\psi_b,m)-E(\psi,m)],
\label{aa6}
\enea
where, $E(\psi,m)$ is incomplete Elliptic integral of second kind. Here
$b=\gamma_m+\sqrt{\gamma_m^2-1}$, $m=2 \sqrt{\gamma_m^2-1}/b$, and 
$y(y_b)$ is related to $\psi(\psi_b)$ as 
\beea
y=\gamma_m+\sqrt{\gamma_m^2-1}\cos(2\psi).
\label{aa7}
\enea
Eq. (\ref{aa5}) and Eq. (\ref{aa6}) together with the above relation  
will give the solution for the electric field behind the proton beam.
 
\end{subsection}

FIG. (\ref{fig.1}) shows the wake wave electric field 
and the perturbed electron density profiles as obtained from the solutions inside as well as outside the beam. 
As mentioned before, length of a proton beam available today is much  longer 
in size than the plasma wavelength. However, due to self
modulation, the beam splits into a train of small proton bunches. Their sizes are 
comparable or even shorter than the typical plasma wavelength. Keeping this in mind, we have  discussed the wake wave excitation by a single proton beam whose size is about three times the plasma wavelength ($\sim 2.8 \lambda_p$). In a later section we discuss the wake field excited by multiple proton bunches. 
From the figure it is seen that maximum electric field amplitude of several MV/m can 
be achieved for plasma densities of n$_0=10^{14}-10^{15}$ cm$^{-3}$ (AWAKE experiment) 
with the beam density $\alpha_0=0.5$ as considered here.

\end{section}

\begin{section}{2D OSIRIS Simulation on PDPWFA }
 
We have carried out our simulation using 2D Particle in Cell (PIC) code OSIRIS.\cite{fon,hem} The simulation shows that a rectangular rigid proton beam with energy $\sim$ 15 GeV interacts with a plasma and forms wake wave. Plasma is uniform in a rectangular box of dimension 53 cm $\times$ 10.62 cm.
Preformed plasma with uniform density $n_0=1.0 \times 10^{14} $ cm$^{-3}$ has been considered. 
 The beam is considered to be rigid which maintains  constant normalized density $\alpha_0=0.5$ throughout the whole simulation run. It is extended 9.3 mm and 5.31 cm in the longitudinal and transverse
direction respectively.
Simulation has been done using a moving window algorithm with window size of 5.31 cm $\times$ 8 cm and resolution of 11 $\mu$m and 80 $\mu$m in the longitudinal and transverse
direction respectively. The number of macro-particle per cell was 16 with total number of $5 \times 10^{6}$ cells in the simulation box.

FIG. (\ref{fig.6}) shows the longitudinal electric field  profile for the excited wake wave. The beam is propagating along positive x-direction and behind the driving beam the wake wave is excited. The perturbed electron density profile is shown in FIG. (\ref{fig.7}). One can easily observe exact matching of the longitudinal field and density profile of our 2D simulations with the 1D exact analytical solution presented in the previous section (FIG.1). This matching is not quite surprising because of the fact that the transverse extension of the beam is much larger compared to its longitudinal size. However, in 1D theory, it was not possible to extract any information of the transverse electric field produced in the beam plasma interaction process.  The transverse field profile  obtained from our 2D simulation is shown in  FIG. (\ref{fig.8}). In the case of side injection of witness electrons, the electrons bunch propagates making
small angle with the driver beam and are gradually sucked in at the proper phase by this transverse electric field.\cite{assmann}

\begin{figure}[h!]
{\centering
{\includegraphics[width=3.0in,height=2.0in]{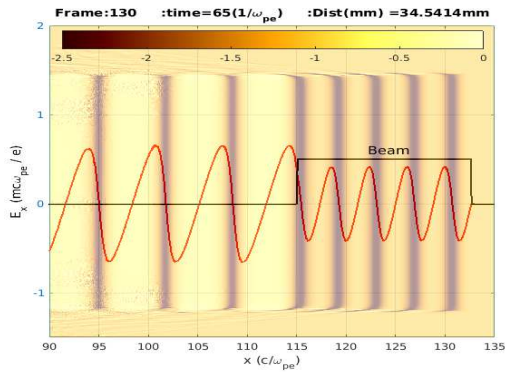}\par}}
\caption{Simulation results: normalized on-axis accelerating electric field and plasma electron density (shown in colour bar) after  
the accelerating proton bunch travels a distance of 34.5 mm in 65 plasma periods.}
\label{fig.6}
\end{figure}

\begin{figure}[h!]
{\centering
{\includegraphics[width=3.0in,height=2.0in]{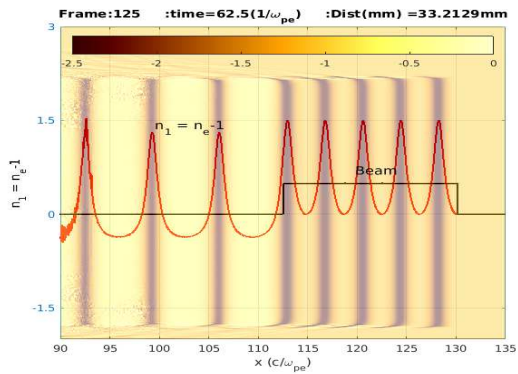}\par}}
\caption{Simulation results: normalized perturbed plasma electron density profile (red curve) after  
the accelerating proton bunch travels a distance of 33.2 mm in 62.5 plasma periods.}
\label{fig.7}
\end{figure}

\begin{figure}[h!]
{\centering
{\includegraphics[width=3.0in,height=2.0in]{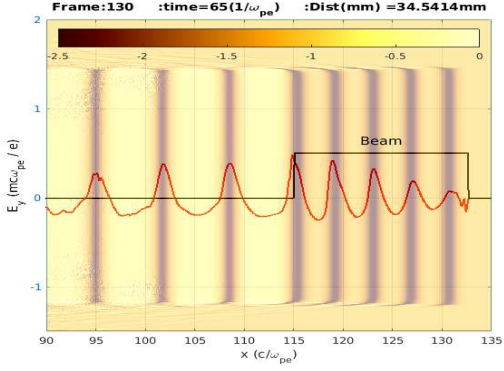}\par}}
\caption{Simulation results: normalized transverse electric field and plasma electron density (shown in colour bar) after the accelerating proton bunch travels a distance of 34.5 mm in 65 plasma periods.}
\label{fig.8}
\end{figure}

\end{section}

\begin{section}{Investigation of PDPWFA using Pseudo potential approach}
Here we present an alternative approach to arrive at the wake wave solution in the 
proton driver scheme. Unlike the solution obtained in the previous section, here 
we relax the assumption of immobility of plasma ions and also consider the velocity of 
the beam to be arbitrary. 
In the process of generation of plasma waves using ultra-relativistic charged particle beam , 
plasma ions carry the main part of the momentum of the source (proton/electron beam). 
In the strong field excited behind the beam, the plasma ions can reach a velocity which is sufficient to make 
a contribution in the process of charge separation and thereby this can 
influence the excited wake field structures.\cite{khachatryan,gorbunov}
Thus, in our investigations we have included the plasma ion motion as well.
We have considered 
the motion of the plasma ions to follow non-relativistic dynamics.

We rewrite the basic set of equations describing proton beam driven nonlinear 1D plasma waves by
incorporating the plasma ion motion as,

\beea
\partial_t{n_j}+ \partial_{x} (n_j v_j)=0, 
\label{wb1}
\enea

\beea
\left(\partial_t+v_j\partial_{x}\right)(\gamma_jv_j)={q_jE}/{m_j},
\label{wb2}
\enea

\beea
\partial_{x}{E}=4\pi \left[\sum_j q_jn_j+e n_b \right].
\label{wb3}
\enea
Here $j$ stands for electrons or ions with $q_j=-e$ for electrons, $q_j=e$ for
ions. The relativistic Lorentz factors associated with electron or ion are denoted by $\gamma_j$.
The proton beam density and the velocity are $n_b$ and $v_b$ respectively with $v_b=v_{ph}$.
All the other variables used here have their usual meanings.

A stationary wave solution is obtained with the same variable transformation 
as has been done in the earlier section.
The plasma electron and ion fluid density both normalized by equilibrium plasma density,
expressed as functions of the electrostatic 
potential $\varphi$ are
\beea
N_e=\beta_{ph}\gamma^2\left[\frac{\varphi_e}{(\varphi_{e}^2-\gamma^{-2})^{1/2}}-\beta_{ph}\right],
\label{wb4}
\enea

\beea
N_i=\frac{\beta_{ph}}{\sqrt{\beta_{ph}^2+2\varphi_i}}.
\label{wb5}
\enea
Here $\varphi_i=-\mu \varphi$ and $\varphi_e=1+\varphi$,  with $m_e/m_i=\mu$ being the electron to ion mass ratio.
Using these expressions for species densities,
from the Poisson's equation, we obtain a second order
differential equation for $\varphi$ as,
%
\begin{figure}[h!]
{\centering
{\includegraphics[width=3.0in,height=2.0in]{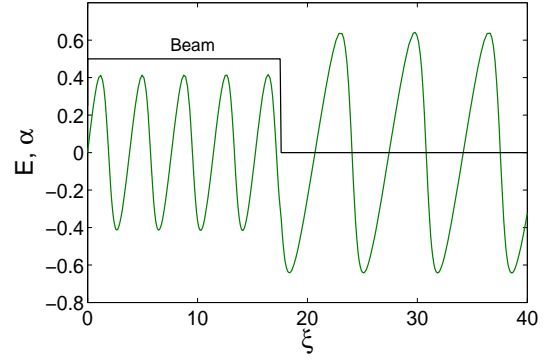}\par}}
\caption{Normalized  wake wave electric field in a proton beam driven 
plasma [$\beta_{ph}=0.995$, $\beta_b=0.995$ and $\mu=1/1836$],
with beam density [$\alpha=0.5$ for $0 \le \xi \le 5.6 \pi$ and zero otherwise].}
\label{fig.2}
\end{figure}

\begin{figure}[h!]
{\centering
{\includegraphics[width=3.0in,height=2.0in]{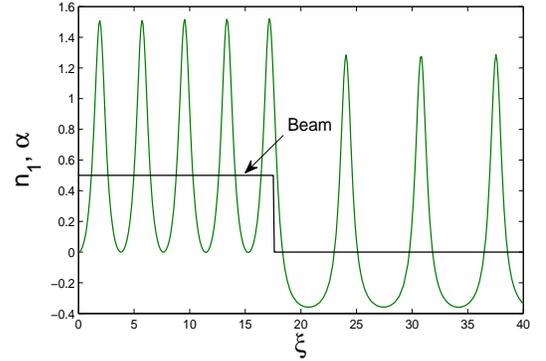}\par}}
\caption{Normalized perturbed electron density in a beam driven plasma
[$\beta_{ph}=0.995$, $\beta_b=0.995$ and $\mu=1/1836$],
with beam density [$\alpha=0.5$ for $0\le \xi \le 5.6 \pi$ and zero otherwise].}
\label{fig.3}
\end{figure}

\begin{equation}
\nder{2}{\varphi}{\xi}=-\fder{U(\varphi)}{\varphi}
\label{wb6}
\end{equation}
where,
\begin{equation}
U(\varphi)=\frac{\beta_{ph}^3}{\sqrt{\beta_{ph}^2+2\varphi_i}}+
\frac{\beta_{ph}^3\gamma^2\varphi_e}{\sqrt{\varphi_{e}^2-\gamma^{-2}}}
-\beta_{ph}^4\gamma^2-\alpha\beta_{ph}^2.
\label{wb7}
\end{equation}
with $\alpha=n_b/n_0$ being the normalized proton beam density. This 
second order differential equation describes the motion of a fictitious particle of 
unit mass in a Pseudopotential $U(\varphi)$. We solve this equation numerically and get 
the solution for the wake wave electric field excited inside and outside a rectangular 
proton beam as well as corresponding perturbed plasma electron density as shown 
in the FIG. (\ref{fig.2}) and FIG. (\ref{fig.3}) respectively. Corresponding to a beam density $n_b=n_0/2$ and 
beam velocity $v_b=0.995c$, the transformer ratio (the ratio of the 
maximum accelerating field behind the beam to the maximum decelerating field inside the beam) which determine 
the overall energy efficiency of the accelerated particles can be 
evaluated easily from this field profile. 

\begin{figure}[h!]
{\centering
{\includegraphics[width=3.0in,height=2.0in]{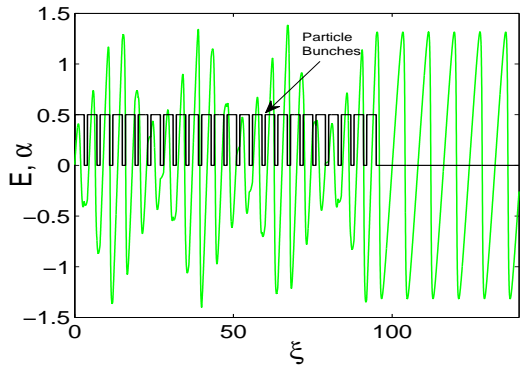}\par}}
\caption{Variation of normalized wake wave electric field in absence of magnetic field driven by a train of 
equidistant particle bunches [$\beta_{ph}=0.995$, $\beta_b=0.995$, and $\mu=1/1836$].}
\label{fig.4}
\end{figure}

\begin{figure}[h!]
{\centering
{\includegraphics[width=3.0in,height=2.0in]{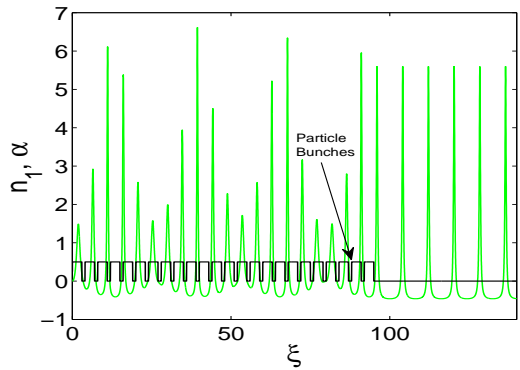}\par}}
\caption{Normalized perturbed electron fluid density in absence of magnetic field driven by a 
train of equidistant particle bunches [$\beta_{ph}=0.995$, $\beta_b=0.995$, and $\mu=1/1836$].}
\label{fig.5}
\end{figure}
 
So far we have discussed the wake wave excited by a single long proton drive beam. However,
it happens that due to self-modulation, the long proton bunch splits 
into a long chain of equi-spaced micro-bunches by the strong focusing and defocusing
forces of the field. So it is of fundamental interest to see how 
strong wake field can be excited behind such multi-beams. FIG. (\ref{fig.4}) and FIG. (\ref{fig.5}) 
represent the electric field and perturbed beam density respectively for the wake wave excited 
by equi-spaced multi proton bunches (length of the each single beam is $\sim 0.5 \lambda_p$ with 
separation distance $ \sim 0.16 \lambda_p$ ) as obtained by numerical solution of Eq. (\ref{wb6}).
From the FIG. (\ref{fig.4}), it is observed that electric field amplitude can not grow indefinitely with 
the increase of the number of beams, rather, the field saturates to a particular limit. The theory mimics the excitation of wake field by self-modulated proton beam. The AWAKE experiment seeks to excite a plasma
wave in this way. It is planned to have an initial plasma stage that micro-bunches a long
proton beam into many, approximately equally spaced bunches and then use this train of bunches in a
second stage to excite a plasma wave and accelerate a secondary electron beam. The above results
could be applicable to the second stage. 


\end{section}


\begin{section}{Conclusion}

We have obtained exact analytical solution for the relativistic wake wave excited by proton beam using fluid approach. The solution, specifically the longitudinal electric field and the density profile, match with the results obtained
from 2D Particle in Cell OSIRIS simulation. The parameters considered here closely resemble the experimental work on PDPWFA. PIC simulation also reveals the characteristics of excited transverse electric field. This field plays a crucial role in trapping the side-injected electrons at proper phase to the excited longitudinal wake wave. Furthermore,
we employed pseudo potential method as an alternative way to deduce the solution for the wake driven by a long chain of equispaced micro-bunches of charge.
The results conclusively shows that the amplitude of the electric field does not grow indefinitely with increase in number of beams. On the contrary, it saturates to a definitive limit.

\end{section}

 \begin{section}{Acknowledgement}
The authors would like to acknowledge the OSIRIS
Consortium, consisting of UCLA and IST (Lisbon, Portugal) for the for
providing access to the OSIRIS 4.0 framework. Work supported by NSF
ACI-1339893.
\end{section}



\end{document}